\documentstyle[floats,aps,epsf]{revtex}
\begin{document}
\title
{New effective nuclear forces with a finite-range three-body 
term and their application to AMD+GCM calculations }

\author{Y. Kanada-En'yo}

\address{Institute of Particle and Nuclear Studies, \\
High Energy Accelerator Research Organization,\\
1-1 Oho, Tsukuba, Ibaraki 305-0801, Japan}

\author{Y. Akaishi}

\address{Institute of Particle and Nuclear Studies, \\
High Energy Accelerator Research Organization,\\
1-1 Oho, Tsukuba, Ibaraki 305-0801, Japan}

\maketitle
\begin{abstract}
We propose new effective inter-nucleon forces with a finite-range
three-body operator.
The proposed forces are suitable 
for describing the nuclear structure properties over a
wide mass number region, including 
the saturation point of nuclear matter.
The forces are applied to microscopic calculations of 
$Z=N$ ($A\le 40$) nuclei and O isotopes  
with a method of antisymmetrized molecular dynamics.
We present the characteristics of the forces and discuss the importance of 
the finite-range three-body term.

\end{abstract}


\section{Introduction}

Experimental information concerning nuclear structure has been 
rapidly increasing in the region of unstable nuclei as well as stable nuclei.
Together with the increase of experimental data,
microscopic calculations of nuclear structure have developed
to describe those structure properties. 
For a systematic study of various types of structures 
covering a wide region of the nuclear chart,
the key issue in microscopic approaches is an extension of the  
model space.
Corresponding to the advancement of microscopic calculations, one must 
improve the effective forces so as to be suitable for 
extended model space.

One of the basic properties of nuclei is saturation of the binding 
energy and density. The saturation property implies 
that a nucleus 
can be divided into fractional nuclei, and that 
such nuclei can fuse into a unified nucleus
with only a small amount of energy.
This ``easiness of assembling and disassembling nucleons''\cite{AMDrev}
 is one of 
the fundamental features of the nuclear many-body system.  
In nuclear-structure problems, this dynamics appears as a variety
of structures such as clustering of nucleons and mean-field formation.
Indeed, it is known that there exist many 
cluster states not only in stable nuclei, but also in unstable nuclei.
This means that the basic dynamics of the nuclear structure  
contains two very different fundamentals. One is the formation of 
clusters, and the other is the formation of a mean field. 
With the traditional microscopic approaches, it was difficult to
describe both of the features
in one framework because of the limitation of the model space. 
Typical mean-field approaches, such as Hartree-Fock(HF) methods, 
are applicable to studying the mean-field aspect in the 
heavy nuclear region, while
they are not suitable for studying cluster structure in light nuclei.
On the other hand, cluster features in light stable nuclei were
studied by cluster models. 
Recently, hybrid models, which can describe both the cluster aspect
and the mean-field aspect, were developed and applied to structure studies
of unstable nuclei as well as stable nuclei.
One of the powerful approaches is a method of antisymmetrized molecular 
dynamics(AMD) \cite{ENYObc,ENYOsup}, 
which is able to describe two different dynamics, i.e. 
the clustering and the mean-field formation
due to its ab initio nature. 
In the study of unstable nuclei in the $p$-shell and $sd$-shell regions, 
performed with the AMD \cite{AMDrev,ENYObc,ENYOsup} method, it was found that
both the cluster features and mean-field are essential 
in a systematic study of the ground and excited states of unstable nuclei 
as well as stable nuclei.
As for an extended model based on 
three-dimensional HF calculations\cite{BONCHE,TAJIMA},
a variation after the parity-projection method\cite{TAKAMI} 
might describe a variety of structures in light nuclei.
In a systematic study with such extended models, 
the quality of the adopted effective interactions becomes
an increasingly serious problem, because the structure is
based on balance and competition between two different dynamics, 
like cluster formation and mean-field formation.
This means that some theoretical results are sensitive to 
the adopted effective force, namely the Hamiltonian, which governs the system.
In other words, the effective force must be appropriate to describe the  
characteristics of various states which may appear
in the extended model space.

In microscopic calculations of nuclear structure,
one often uses the effective forces to solve nuclear many-body problems.
In most of the phenomenological 
effective inter-nucleon forces,
the dominant part is expressed by central forces.
With the Minnesota force \cite{MINNESOTA}
and the Volkov force \cite{VOLKOV}, which consist of two-body central forces, 
the properties of very light nuclei are well reproduced.
However, it is impossible to describe the saturation property of
 nuclear matter and heavy nuclei (mass number $A \ge 10$)
with such two-body central forces.
For satisfying the saturation properties, 
phenomenological density-dependent repulsive terms 
are usually imported in addition to the central two-body terms.
Such forces with a density dependence are Gogny forces\cite{GOGNY}, 
Skyrme-type forces\cite{CHABANAT,SIII}
and Modified Volkov forces\cite{MVOLKOV}, for example. 
These forces contain zero-range
density-dependent operators or zero-range three-body operators.
HF calculations with Gogny forces and Skyrme forces have succeeded in 
describing the
bulk properties of the low-lying states of heavy nuclei. 
These forces, however, fail to reproduce the properties 
of very light nuclei.
Tohsaki introduced repulsive terms with finite-range three-body operators,
and proposed new effective forces responsible for the microscopic
$\alpha$-cluster model, which succeeded to describe the 
overall properties over a wide range of masses from $\alpha$ to nuclear matter
\cite{TOHSAKInew}.
Namely, the $\alpha$-$\alpha$ scattering behavior, 
the size and energy of double closed shell nuclei($\alpha$, $^{16}$O and 
$^{40}$Ca) and the matter
saturation property were successfully reproduced with Tohsaki's forces. 
However, unfortunately,
the forces contain too complicated three-body operators 
to be employed in practical microscopic calculations of 
general nuclei.

Our aim in this paper is to propose new effective inter-nucleon 
forces which are applicable
to practical calculations of general nuclei with the AMD method.
In order to save computational costs,
we introduce a simple three-body operator expressed by a one-range Gaussian.
The new forces are applied to microscopic calculations of
$Z=N$ nuclei up to $^{40}$Ca and 
neutron-rich O isotopes with the AMD method.
We explain the characteristics of the proposed forces and prove the
success in describing the overall properties from the $\alpha$ particle 
to nuclear matter, by demonstrating
the results of the radii and energy of finite nuclei,
the elastic $\alpha$-$\alpha$ scattering phase shift, 
and the saturation property of nuclear matter. 
We point out the significance of finite-range 
three-body terms in the effective forces
from a phenomenological point of view, and discuss the
role of the repulsive three-body terms in the connection 
with the tensor force, which is the origin of matter saturation.

This paper is organized as follows.
The new effective forces with a finite-range three-body term are proposed
in \ref{sec:central}.
We describe the characteristics of the effective forces
by demonstrating the calculated results with the new forces
in \ref{sec:results}.
In \ref{sec:discuss},
the significant role of the finite-range
three-body term is discussed.
Finally, in \ref{sec:summary} we summarize the present work.

\section{Effective central forces}
\label{sec:central}

As mentioned above, 
for most of the phenomenological effective forces in microscopic 
models, the dominant part is the central force.
Concerning the effective central forces,
it is regarded as a tensor force, 
a hard core and other terms in the bare interactions are 
renormalized because it is not easy to
explicitly treat such terms in microscopic models
for a nuclear many-body system. 
By categorizing the effective central forces into three types,
we briefly review the general 
tendency of the typical effective forces to reproduce the
structure properties, such as the saturation behaviour
of nuclei and nuclear matter and the $\alpha$+$\alpha$
inter-cluster potential.

(1)
Regarding finite-range two-body forces with no density-dependent terms,
there exist such effective forces as Minnesota\cite{MINNESOTA} and 
Volkov\cite{VOLKOV} forces
which can reproduce the $\alpha$+$\alpha$ phase shift and 
the size and binding energy of an $\alpha$ particle
and light $p$-shell nuclei. However it is impossible to
represent the saturation of matter and heavy nuclei
with such the two-body central forces.

(2)
Gogny\cite{GOGNY} forces, 
Skyrme-type\cite{SIII,CHABANAT} forces and 
Modified Volkov\cite{MVOLKOV} forces,
which can guarantee nuclear matter saturation,
contain zero-range density-dependent terms or zero-range three-body
terms.
Gogny forces and Skyrme-type forces successfully explain the size and
the binding energy of the overall heavy nuclei, and also matter saturation.
In spite of the success of nuclear matter saturation, 
all of these effective forces 
fail to reproduce the size of the $\alpha$ particle and the $\alpha$-$\alpha$ 
scattering behaviour.
Namely, all of these forces give larger $\alpha$-particle sizes
and a less $\alpha$-$\alpha$ inter-cluster potential than the 
experimental ones.
Therefore, the applicability of such forces to 
very light nuclei is not assured.

(3)
We know only one kind of the effective force which can simultaneously 
guarantee the size of $\alpha$ particle, the $\alpha$-$\alpha$ phase shift, and
the saturation properties of nuclei and nuclear matter. 
That is the third type of effective force containing 
finite-range three-body operators,
proposed by Tohsaki \cite{TOHSAKInew}, namely, 
Tohsaki's F1 and F2 forces.
It was found that these effective forces can successfully reproduce 
the $\alpha$-$\alpha$ phase shift, the sizes and binding energy of 
double-closed
nuclei up to $^{40}$Ca and also matter saturation point.

In the present paper, we propose new effective forces which 
can be classified into the last category (3),
forces with finite-range three-body operators.
To reproduce the nuclear matter saturation properties,
the effective central forces must contain density-dependent terms.
It is considered that the origin of the density-dependent terms 
is suppression of the tensor force of bare nucleon-nucleon interactions
 in the nuclear matter.
In other words, the density-dependent terms are considered to
phenomenologically simulate the medium effect that 
tensor force is suppressed in the nuclear matter because of the Pauli principal.
From this point of view, it seems to be
reasonable that the effective central forces contain finite-range
three-body terms because the original tensor force is the middle-range one. 
Moreover, the type (2) forces with 
zero-range density-dependent terms have a fatal difficulty in 
simultaneously describing the density and the energy of an $\alpha$ particle 
and nuclear matter. This failure in the quantitative reproduction of the 
saturation properties with type (2) forces is proved later in
\ref{sec:discuss}.

We propose effective central forces, $V_{central},$ 
expressed by two-body and three-body
operators as follows:
\begin{equation}\label{eqn:f3b}
V_{central}=\sum_{i<j}V^{(2)}_{ij}
+\sum_{i<j<k}V^{(3)}_{ijk},
\end{equation}
where
\begin{eqnarray}
&\label{eqn:f3b-2}
V^{(2)}_{ij}=(1-m+b P_\sigma - h P_\tau -m P_\sigma P_\tau)
\sum_{n=1}^2 
\left\lbrace
v^{(2)}_n \exp\left[-(\frac{{\bf r}_i-{\bf r}_j}{a^{(2)}_n})^2 \right] 
\right\rbrace + 
v^{(2)}_3 \exp\left[-(\frac{{\bf r}_i-{\bf r}_j}{a^{(2)}_3})^2 \right] 
 \\
&
V^{(3)}_{ijk}=v^{(3)} \exp\left[-\frac{1}{{a^{(3)}}}
\left\lbrace ({\bf r}_i-{\bf r}_j)^2
+({\bf r}_j-{\bf r}_k)^2
+({\bf r}_k-{\bf r}_i)^2 \right\rbrace
\right].
\\
\end{eqnarray}
Here we adopt a simple operator for the finite-range three-body terms.
Compared with Tohsaki's forces expressed by three-range three-body operators, 
this one-range three-body operator  with a symmetric form in the 
present forces is suitable for practical microscopic calculations 
of general nuclei with the AMD method.  

We propose a set of parameters for the ranges and strengths 
of the two-body and three-body terms. The dominant components of the two-body
terms are the first term in Eq.\ref{eqn:f3b-2}.
We adopt the same form as the Volkov force for this term, which consists
of a medium-range attractive Gaussian and a short-range repulsive Gaussian.
Wigner, Bertlett, Heisenberg and Majorana terms are chosen to 
be same parameters as the common multiplier of these two range Gaussians.
The basic idea of the derivation of the parameters is that
the nuclear structure properties over a wide mass number region should be
described with a two-range two-body force and an one-range three-body force
without mass-number depending parameters.
We adopt the same range parameters for the two-range Gaussian of the 
two-body force as the Gogny force($a_1^{(2)}=1.2$(fm) and $a_2^{(2)}=0.7$(fm)).
With longer range parameters, we failed to obtain sufficient reproduction 
of the nuclear structure. Although these range parameters for the 
two-body terms are shorter than the Volkov and MV1 forces, 
the range of the inter-cluster potential is not so different 
from those given by the MV1 forces, as shown in later, 
due to the effect of the repulsive finite-range three-body term in the present
force.
In order to explain the nucleon-nucleon scattering phase shift,
the form with a two-range two-body and an one-range three-body is 
too simple because of the restriction of the simplified three-body operator.
We add a small amount of a repulsive two-body term given by the second term
of the two-body force in eq.\ref{eqn:f3b-2},
$v^{(2)}_3 \exp\left[-({\bf r}_i-{\bf r}_j)^2/{a^{(2)}_3}^2 \right]$, 
and tune again the strength of this term and the three-body term.
In table \ref{tab:f3b}, the proposed parameter sets of the present forces 
are presented.
We call the present parametrization as the F3B forces.
Two sets case(1) and case(2) of parameters of $m$, $b$ and $h$ for Wigner, 
Bertlett, Heisenberg and Majorana terms are listed in table \ref{tab:f3b}.
Comparing the case(1) and case(2) parameters, the triplet and singlet even 
components are same but the odd components different between two
sets (1) and (2).
As a result, the case(1) and case(2) interactions give the same results for
double closed $Z=N$ nuclei written by HO wave functions, 
the $\alpha$-$\alpha$ interaction, and total energy of symmetric 
nuclear matter, however, they give different results for $Z \ne N$ nuclei.

The present parameters $m$ for the Majorana term shown in table \ref{tab:f3b} 
are determined so as to reproduce the binding energy of $Z=N$ nuclei 
in mass region $A=4\sim 40$ with the present AMD+GCM calculations.
The Majorana parameter $m$ is considered to be an adjusting parameter 
in fine tuning to fit the experimental data.
It means that one can modify the value $m$, which is suitable 
to his model wave function.

In practical AMD calculations of $Z=N$ nuclei and
O isotopes, we also employ the two-body spin-orbit force in addition to these
central F3B forces.

\begin{table}
\caption{ \label{tab:f3b} Parameter sets of the effective 
central force(F3B force) in Eq. \protect\ref{eqn:f3b}. }
\begin{center}
\begin{tabular}{cccc|cc|cc|c}
 $v_1^{(2)}$(MeV) &$v_2^{(2)}$(MeV) & $a_1^{(2)}$(fm) & $a_2^{(2)}$(fm) & 
$v_3^{(2)}$(MeV) &  $a_3^{(2)}$(fm) & $v^{(3)}$(MeV) & $a^{(3)}$(fm$^2$) &
$\begin{array}{cccc}
\qquad  & \ \ \  m\ \ & \ \ \ b\ \ & \ \ \ h\ \ \end{array}$\\
\hline
  $-198.34$ & 300.86 &  1.2 & 0.7  & 22.5 & 
0.9  &600  & 
 1.25 & 
$\begin{array}{cccc}
 {\rm case(1)}& 0.421 & 0.1 & 0.085 \\ 
{\rm case(2)}& 0.193 & -0.185 & 0.37\end{array}$ \\ 
\end{tabular}
\end{center}
\end{table}

\section{Characteristics of the effective forces}
\label{sec:results}

In this section, we exhibit the
characteristics of the proposed effective 
forces concerning the saturation properties and $\alpha$-$\alpha$ 
scattering behavior 
compared with other types of typical phenomenological 
effective forces. 

\subsection{$\alpha$-$\alpha$ interaction}

The $\alpha$-$\alpha$ inter-cluster interaction is important 
for studying the cluster structure of light nuclei.
The scattering phase shift is a useful physical quantity
to inspect the features of the inter-cluster interaction. 
The phase shift of the elastic $\alpha$-$\alpha$ scattering is calculated
by the resonating group method(RGM), where
an $\alpha$ particle is simply written using a $(0s)^4$ harmonic oscillator 
wave function \cite{KAMIMURArgm}.
The width parameter of $\alpha$ clusters
is chosen to be the optimum value for an isolate $\alpha$-particle for
each effective force.

\begin{figure}
\noindent
\epsfxsize=0.4\textwidth
\centerline{\epsffile{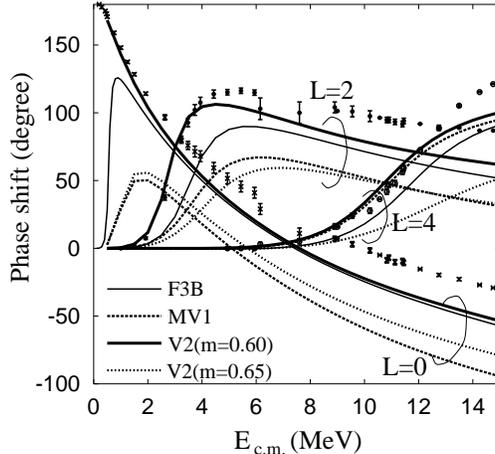}}
\caption{\label{fig:rgm}
Calculated phase shift of elastic $\alpha$-$\alpha$ scattering
with the F3B force, the MV1 case 1($m=0.60$), and the 
Volkov No.2 ($m=0.60$ and $m=0.65$).  
The width parameter($\nu$) for an $\alpha$ particle 
in an RGM calculation was chosen to be 0.257, 0.209, 0.267 (fm$^{-2}$) for 
the F3B, MV1, Volkov No.2 forces, respectively.
The points indicate the experimental data taken from \protect\cite{aa-phase}.}
\end{figure}

In Fig. \ref{fig:rgm},
the calculated $\alpha$-$\alpha$ phase-shift values are compared with the 
experimental data.
It has been known that the experimental phase-shift data
are well reproduced by the Minnesota force\cite{MINNESOTA} and 
the Volkov No.2 force($m=0.60$)\cite{OKABE-be9}.
The theoretical values calculated with the new force(F3B) 
agree reasonably  well with the experimental data.
On the other hand, the MV1 case(1) force($m=0.60$)
underestimates the experimental data, which means that 
the force provides a less-attractive $\alpha$-$\alpha$ interaction. 
It was also found that the 
SII force(a Skyrme-type force) gives a less-attractive inter-$\alpha$ 
interaction, and
therefore, fails to reproduce the $\alpha$-$\alpha$ phase shift
\cite{TOHSAKInew}.

\begin{figure}
\noindent
\epsfxsize=0.4\textwidth
\centerline{\epsffile{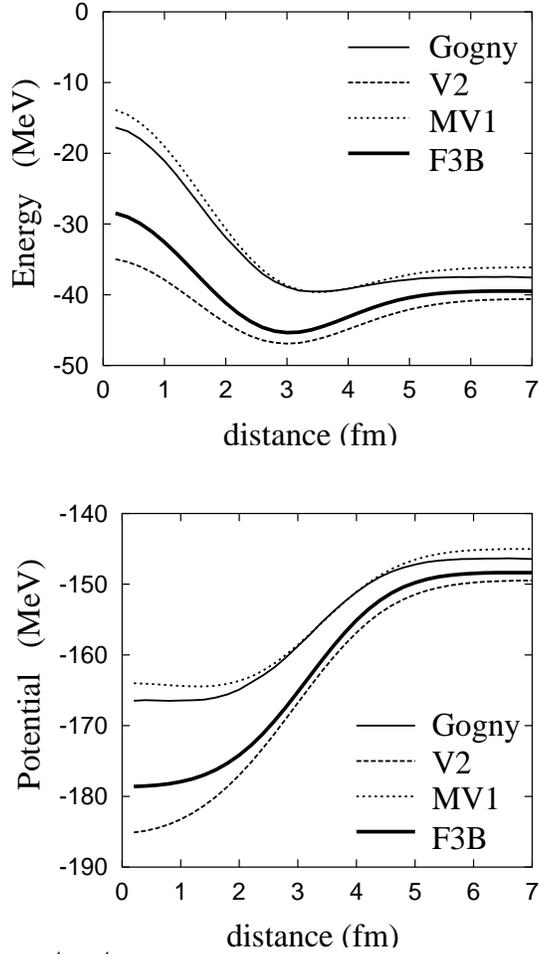}}
\caption{\label{fig:he4-he4}
Energy and potential between $^4$He-$^4$He.
The potential energy(total energy) of the $^4$He-$^4$He system 
as functions of the relative distance 
($d$) is shown in the lower(upper) panel.
The effective forces are Gogny D1, Volkov($m=0.60$), MV1 cases 1($m=0.60$) and
F3B.}
\end{figure}

Figure \ref{fig:he4-he4} presents the energy and potential 
as a function of inter-cluster distance in the
$\alpha$-$\alpha$ system.
The $\alpha$-$\alpha$ system is represented by the Brink-type wave 
function. Namely, 
two $\alpha$ clusters are located at points, $(x,y,z)=(0,0,d/2)$
and $(x,y,z)=(0,0,-d/2)$, where $d$ indicates the inter-cluster distance.
The width parameter of an $\alpha$ is fixed to be $\nu=0.25$ fm$^{-2}$.
All of the energy curves have minimum points at  around $d\sim 3$ fm.  
The energy and potential for F3B force and for 
Volkov No.2($m=0.60$) are very similar to each other 
in the region $d > 2.5$ fm.
In the small-distance region $d<2$ fm, the $\alpha$-$\alpha$ potential 
in the case of the F3B force is shallower
than that for Volkov No.2 because of the three-body repulsive term.
The $\alpha$-$\alpha$ potential for the
Gogny D1 force and the MV1 case(1) force($m=0.60$) are similar to each other.
Both forces indicate a smaller inter-cluster potential 
compared with the Volkov No.2 ($m=0.60$) force. This result 
is consistent with the 
$\alpha$-$\alpha$ scattering behavior.

\subsection{$^{16}$O-$\alpha$ system}

\begin{figure}
\noindent
\epsfxsize=0.4\textwidth
\centerline{\epsffile{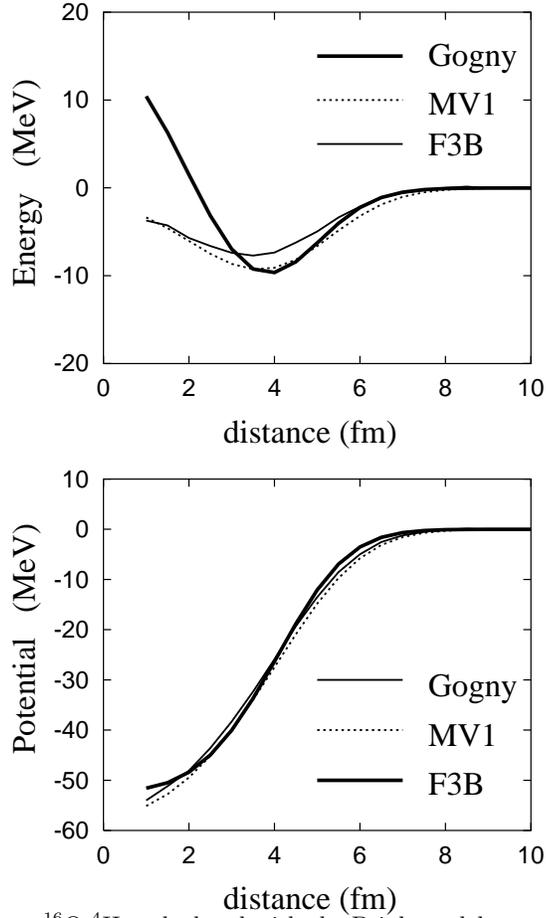}}
\caption{\label{fig:o16-he4-e}
Energy and potential between $^{16}$O-$^4$He calculated with 
the Brink model wave functions.
The potential energy(total energy) of the $^{16}$O-$^4$He system 
as functions of the relative distance 
$d$ is shown in the lower(upper) panel.
The effective forces are Gogny D1, MV1 case 1($m=0.60$) and
F3B forces. Width parameters are chosen to be optimum widths which give
the minimum energy for $^{20}$Ne, i.e. $\nu=0.16$,$\nu=0.165$ and $\nu=0.20$
for the Gogny D1, MV1 case1($m=0.60$) and F3B forces, respectively.}
\end{figure}

\begin{figure}
\noindent
\epsfxsize=0.4\textwidth
\centerline{\epsffile{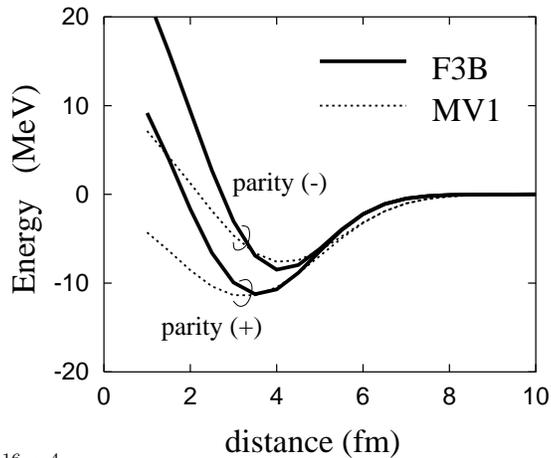}}
\caption{\label{fig:o16-he4-np}
The total energy of the $^{16}$O-$^4$He system for the parity-projected
states calculated with the Brink model wave functions.
$d$ is the relative distance between $^{16}$O and $^4$He.
Width parameters are chosen to be optimum widths which give
the minimum energy for $^{20}$Ne, i.e. $\nu=0.165$ and $\nu=0.20$
for the MV1 case1($m=0.60$) and F3B forces, respectively.}
\end{figure}

\begin{table}
\caption{ \label{tab:ne20} The binding energy of the band head 
states($0^+_1$ and $1^-_1$) of the parity doublets in 
$^{20}$Ne calculated with the AMD+GCM
method.
The adopted Hamiltonian is same as Eq.\protect\ref{eq:hamiltonian}.
The F3B case(1) and the 
MV3 case 1($m=0.60$) are used as the central forces.
The width parameter is chosen to be $\nu=0.20$ and $\nu=0.165$, 
which minimise the binding energy of $^{20}$Ne, for the F3B case(1) 
and MV3 case 1($m=0.60$) forces, respectively.
Excitation energy of the $1^-_1$ states are also shown.}
\begin{center}
\begin{tabular}{c|ccc}
 (MeV)    &  F3B  & MV3 & exp.\\
$0^+_1$ & 157.2 & 153.1 & 160.647 \\
$1^-_1$ & 151.4 & 146.5 &  154.859\\
E$_x$($1^-_1$) & 5.8 & 6.5 & 5.788\\
\end{tabular}
\end{center}
\end{table}

The total and potential energy of the $^{16}$O-$\alpha$ system is 
shown in Fig. \ref{fig:o16-he4-e}
and Fig. \ref{fig:o16-he4-np}. 
The $^{16}$O-$\alpha$ system is 
represented by the Brink-type wave function. Namely, $^{16}$O and $\alpha$
described by the harmonic oscillator wave functions are located at 
$(x,y,z)=(0,0,d/5)$ and $(x,y,z)=(0,0,-4d/5)$, where $d$ is 
the inter-cluster distance. 
The intrinsic energy of clusters is subtracted.
The width parameter of $^{16}$O and $\alpha$
is chosen to be the optimum width which gives
the minimum energy of $^{20}$Ne for each forces.
As shown in Fig. \ref{fig:o16-he4-e}, the potential energy is 
similar among three forces(Gogny, MV1 case 1($m=0.60$) and F3B).
In the total energy (Fig.\ref{fig:o16-he4-e}), 
the minimum energy points exist at almost the same
distance $d\sim 4$ fm for the MV1 case 1($m=0.60$) and F3B forces.
In the small distance region, 
the F3B force has a more repulsive term of the total energy 
than the MV1 case 1($m=0.60$)
force because the width parameter $nu$ is larger for the F3B force
than the MV1 case 1($m=0.60$) force,
therefore, the kinetic energy increases more rapidly as the inter-cluster
distance become small.

The $^{16}$O-$\alpha$ potential energy is closely related to the 
structure of parity doublet states in $^{20}$Ne. 
The $K=0^+_1$ and the $K=0^-_1$ bands in $^{20}$Ne are known 
to be the parity doublet states, which are considered to originate from 
the $^{16}$O+$\alpha$ cluster structure. In Fig. \ref{fig:o16-he4-np}, 
the total energy of the parity-projected states for 
the Brink-type wave functions are plotted as a function of the inter-cluster
distance $d$. Comparing the results with the F3B force and MV1 force, 
the minimum energy is qualitatively similar. The minimum energy point
for the positive-parity states shifts to the large distance region in case 
of the F3B force, which has an effect on the enhancement of the 
$^{16}$O+$\alpha$ clustering in $^{20}$Ne. 

In the above-mentioned discussion, the $^{16}$O-$\alpha$ potential 
are studied within the framework, where the $^{16}$O and $\alpha$ are 
treated as inert clusters written by harmonic oscillator wave functions.
However, in the real $^{20}$Ne, the core excitation is more important 
than the case of $\alpha$-$\alpha$ system. Moreover, the
harmonic oscillator wave function is too simple to describe the real $^{16}$O
nucleus. In order to see agreements to the experimental data, 
we calculate the band-head states($0^+_1$ and $1^-_1$) 
of the parity doublets in $^{20}$Ne with the AMD+GCM method, where
the effect of core excitation is automatically included.
The detailed description of the method is given later. 
As shown in Table \ref{tab:ne20}, the binding energy and the excitation 
energy of the $1^-_1$ state are well reproduced by using the F3B force.

\subsection{Radii and Binding energy of $Z=N$($A\le 40$) nuclei and O isotopes}

In order to discuss the features of the saturation properties 
(radii and binding energy)
in finite nuclei, 
we apply the F3B forces to microscopic calculations based on a AMD method
for $Z=N$($A\le 40$) nuclei and O isotopes.$\nu=0.16$,
The formulation of AMD 
for a nuclear structure study of ground and excited states
is explained in \cite{AMDrev,ENYObc,ENYOe}.
Here, we briefly explain the framework in the present work, namely, 
the generator coordinate method in the framework of AMD with constraint
(AMD+GCM).
The wave function of a system is written by superposition of 
AMD wave functions,
\begin{equation}
\Phi=c \Phi_{AMD} +c' \Phi '_{AMD} + \cdots .
\end{equation}
An AMD wave function of a nucleus with mass number $A$
is a Slater determinant of Gaussian wave packets:
\begin{eqnarray}
&\Phi_{AMD}({\bf Z})=\frac{1}{\sqrt{A!}}
{\cal A}\{\varphi_1,\varphi_2,\cdots,\varphi_A\},\\
&\varphi_i=\phi_{{\bf X}_i}\chi_{\xi_i}\chi_{\tau_i} :\left\lbrace
\begin{array}{l}
\phi_{{\bf X}_i}({\bf r}_j) \propto
\exp
[-\nu({\bf r}_j-{{{\bf X}_i}\over{\sqrt{\nu}}})^2],\\
\chi_{\xi_i}=
\left(\begin{array}{l}
{1\over 2}+\xi_{i}\\
{1\over 2}-\xi_{i}
\end{array}\right),
\end{array}\right. 
\end{eqnarray}
where the $i$th single-particle wave function ($\varphi_i$)
is a product of the spatial wave function ($\phi_{{\rm X}_i}$),
 the intrinsic spin function ($\chi_{\xi_i}$) and 
the iso-spin function ($\chi_{\tau_i}$). 
The spatial part ($\phi_{{\rm X}_i}$) is represented by 
complex variational parameters(${\rm X}_{1i}$, ${\rm X}_{2i}$, 
${\rm X}_{3i}$).
$\chi_{\xi_i}$ is the intrinsic spin function, defined by
$\xi_{i}$, and $\tau_i$ is an iso-spin
function, which was fixed to be up(proton) or down(neutron)
in the present calculations.
Thus, an AMD wave function is expressed by a set of variational parameters,
${\bf Z}\equiv \{{\rm X}_{ni},\xi_i\}\ (n=1,2,3\ \hbox{and }  i=1,\cdots,A)$, 
which stand for the centers of Gaussians of the spatial part 
and the intrinsic spin orientations of the single-particle
wave functions.
By using the frictional cooling method, we performed energy variation 
with respect to a parity-projected AMD wave function, $\Phi^{\pm}$, 
while satisfying the condition that the 
principal oscillator quantum
number of the system, $\langle a^\dagger a\rangle \equiv \langle
\Phi^{\pm}|\sum_{_i}^{A}\hat{{\bf a}}_i^{\dagger}\cdot\hat{\bf
a}_i|\Phi^{\pm}\rangle/\langle\Phi^{\pm}|\Phi^{\pm}\rangle$,
equals a given number.
After variation with the constraint on the oscillator quanta,
we superposed the spin-parity eigen states projected from the 
obtained AMD wave functions so as to diagonalize the Hamiltonian matrix and
the norm matrix.
The expectation values for 
such structure properties as the binding energy and the root-mean-square radius
were evaluated by the superpositions.
In the present calculations, we superposed several spin-parity projected 
AMD wave functions with respect to 
the generator coordinate, $\langle a^\dagger a\rangle$.

The used Hamiltonian in the AMD+GCM calculations was,
\begin{equation}\label{eq:hamiltonian}
{\cal H}=-\frac{\hbar^2}{2M}\sum_i \nabla_i^2-T_G+V_{coulomb}+
V_{central}+V_{spin-orbit},
\end{equation}
where the first term stands for the kinetic energy operator, the second
is for extraction of the center of mass energy, 
and $V_{coulomb}$ means the Coulomb energy term.
The central force, $V_{central}$, is the F3B force.
In addition to the central force, we employ the
following two-body spin-orbit force($V_{spin-orbit}$) :
\begin{eqnarray}
&V_{spin-orbit}=\sum_{i<j} V^{(LS)}, \\
\label{eq:lsforce}& V^{(LS)}_{ij}=\frac{1+P_\sigma}{2} \frac{1+P_\tau}{2} {\bf l}_{ij}\cdot 
{\bf s}_{ij} 
\sum_{n=1}^2 \left\lbrace
v^{(LS)}_n \exp\left[-(\frac{{\bf r}_i-{\bf r}_j}{a^{(LS)}_n})^2 \right] 
\right\rbrace,
\end{eqnarray}
where the same range of parameters as
those of the G3RS force \cite{G3RS}, $a_1^{(LS)}=0.4472$ fm and 
$a_2^{(LS)}=0.5774$ fm, are used.
The strengths of the spin-orbit force were chosen to be
$v^{(LS)}_1=-v^{(LS)}_2=1800$ MeV in the present calculations. 

In Table \ref{tab:radii}, the results of 
root-mean-square-radii(r.m.s.r) 
for point-like proton distributions
of double-closed nuclei, $^4$He, $^{16}$O and $^{40}$Ca
are listed. It is found that, overall, the F3B force well reproduces 
the radii and energy of these nuclei as well as Tohsaki's F1 force. 
It has been already known that it is impossible to systematically describe 
the nuclear saturation properties by Volkov forces. 
For instance, in the case of Volkov No.2(V2) with the Majorana exchange 
parameter $m=0.65$, the radii of $^{16}$O and $^{40}$Ca are 
largely underestimated.
On the other hand, the MV1 force and the SIII force fail to qualitatively 
reproduce the compact size of a real $^4$He. 
A HF+BCS calculation with SIII gives an 
extremely large radius of an $\alpha$ particle 
because the center-of-mass motion is not exactly extracted in the model space.
By an AMD calculation, where the center-of-mass motion 
was microscopically extracted, the calculated $\alpha$ radius 
with the SIII force was found to be 1.66 fm, which is as large as those 
with the MV1 force and the SIII force.
Therefore, we can conclude that 
these forces with zero-range density-dependent 
terms always overestimate the correct size of an $\alpha$ particle.

\begin{table}
\caption{ \label{tab:radii} Calculated binding energy and 
root-mean-square radii for a point-like proton distribution. 
Theoretical values for the F3B case (1) were obtained by AMD+GCM calculations.
The values with the Volkov force No.2(V2),
A modified Volkov force(MV1) case 1 and Tohsaki's F1 force are those 
calculated by the closed-shell HO wave functions 
(\protect\cite{MVOLKOV,TOHSAKInew}).
The center-of-mass correction 
in the theoretical radii for V2, MV1, Tohsaki's F1 forces is subtracted as 
$\langle r^2 \rangle - 3/2a A$ ($A$ is the mass number and $a$ is 
the harmonic oscillator size parameter related to 
$\hbar\omega=\hbar^2 a/M$, where $M$ is the  nucleon mass.)
The theoretical values of the Skyrme III force are three-dimensional HF+BCS
calculations by N. Tajima et al \protect\cite{TAJIMA}.
The value in parenthesis was calculated with a single 
AMD wave function (the values for SIII are taken from 
\protect\cite{SUGAWA}).
The experimental r.m.s.r. were deduced from the charge 
radii\protect\cite{NDATA}.}
\begin{center}
\begin{tabular}{cccccccc}
& Force  & V2($m=0.65$) & MV1(case1) & SIII & Tohsaki's F1 & F3B(this work)
& exp.\\ 
\hline
$^4$He & $E$(MeV) & 28.0 & 28.2 & 26.5(25.6) & 27.5 & 28.2(28.0) & 28.296   \\
 &  r.m.s.r.(fm) & 1.46 & 1.64 & 1.88(1.66) & 1.50 & 1.52(1.48) & 1.47   \\
$^{16}$O & $E$(MeV) &  105.9 & 122.2 & 128.1(120.9) & 123.0 &129.7 &127.617 \\
 &  r.m.s.r.(fm) & 2.35 & 2.62 & 2.63(2.60) & 2.48 & 2.57 & 2.57 \\
$^{40}$Ca & $E$(MeV) & 348 & 351.6 & 341.8 & 334.0 & 350.6 & 342.046  \\
 &  r.m.s.r.(fm) & 2.64 & 3.31 & 3.40 & 3.38 & 3.38 & 3.39 \\
\end{tabular}
\end{center}
\end{table}

\begin{figure}
\noindent
\epsfxsize=0.4\textwidth
\centerline{\epsffile{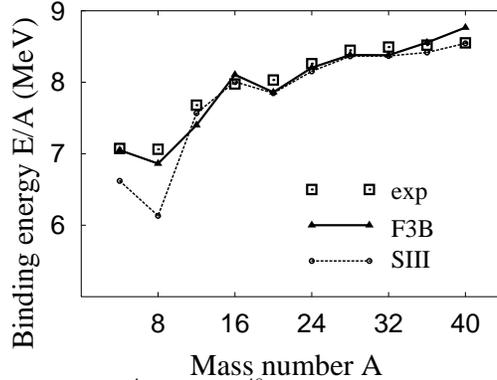}}
\caption{\label{fig:be}
Binding energy of $Z=N$ nuclei from $^4$He up to $^{40}$Ca,
calculated by AMD+GCM with the F3B case (1) compared with the experimental data.
The theoretical values by HF+BCS calculations 
with SIII \protect\cite{TAJIMA} are also displayed.}
\end{figure}

\begin{figure}
\noindent
\epsfxsize=0.4\textwidth
\centerline{\epsffile{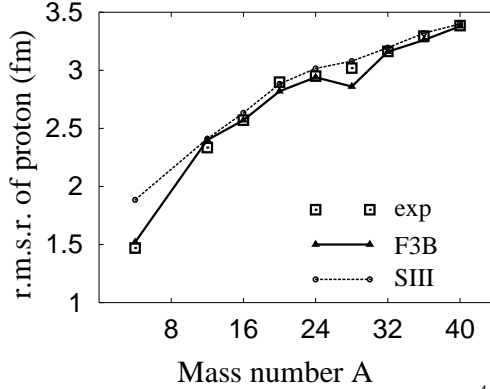}}
\caption{\label{fig:radii}
Root-mean-square radii of the proton density for 
$Z=N$ nuclei from $^4$He up to $^{40}$Ca 
calculated by AMD+GCM with the F3B case (1).
The experimental data were deduced from the
charge r.m.s.r.\protect\cite{NDATA}.
The theoretical values by HF+BCS 
with SIII\protect\cite{TAJIMA} are also shown.}
\end{figure}

In Figs. \ref{fig:be} and \ref{fig:radii},
the present results of the binding energy and the proton r.m.s.r. 
of $Z=N$ nuclei up to $^{40}$Ca 
with the F3B forces are compared with the experimental data.
The present results well agree with the overall data in 
this nuclear region. 

\begin{figure}
\noindent
\epsfxsize=0.4\textwidth
\centerline{\epsffile{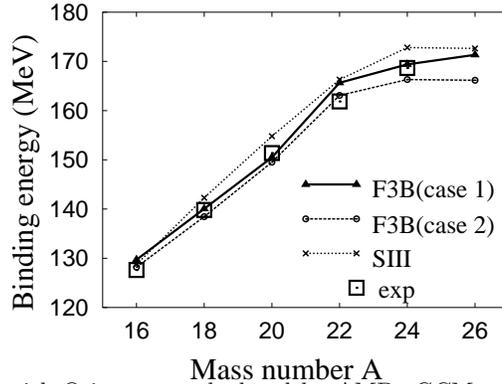}}
\caption{\label{fig:obe} 
Binding energy of neutron-rich O isotopes 
calculated by AMD+GCM with F3B case 1 and case 2.
The HF+BCS calculations with SIII\protect\cite{TAJIMA} and 
the experimental data are also displayed.}
\end{figure}

Although the parameter sets, case (1) and case (2), of the F3B force
give almost the same results for $Z=N$ nuclei, the results 
for $Z\ne N$ nuclei are different
because the ratios of the inter-nucleon interaction for the isospin $T=1$ 
system to that for the $T=0$ system
are different between cases (1) and (2).
The theoretical binding energy of neutron-rich 
O isotopes are compared with the experimental data in 
Fig. \ref{fig:obe}. 
The calculated results with both sets of interaction parameters
agree reasonably with the experimental data.
$^{26}$O is known to be a particle-unstable nucleus. 
Concerning the systematics of 
the two-neutron separation energy,
the neutron drip line in O isotopes 
is reproduced by the case (2) force. Although 
the drip line is not described by the case (1) force, 
we can not conclude that the parameter case (1) is inappropriate at the 
present stage
because the present model space seems to be insufficient 
to describe the detailed structure of $^{24}$O, as conjectured from a 
disagreement between the theoretical radius and the experimental one.

\begin{figure}
\noindent
\epsfxsize=0.4\textwidth
\centerline{\epsffile{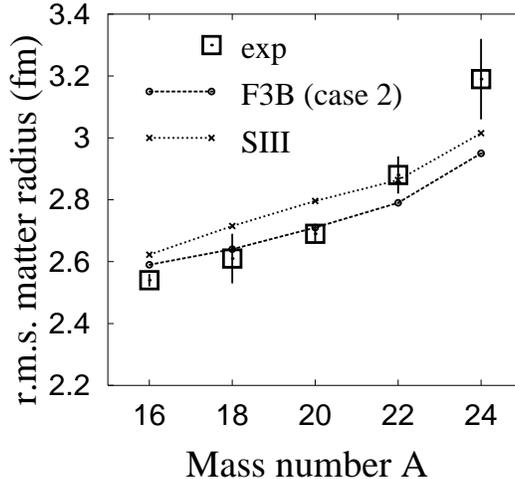}}
\caption{\label{fig:ormsr}
Root-mean-square(r.m.s.) matter radii calculated by AMD+GCM with  
F3B case 2. The experimental data are those deduced from the  
reaction cross sections taken from Ref.\protect\cite{OZAWA}. 
The theoretical values by HF+BCS 
with SIII\protect\cite{TAJIMA} are also shown. 
}
\end{figure}

The present results concerning the r.m.s.r of  O isotopes
quantitatively agree with the experimental data
deduced from the interaction cross-section data\cite{OZAWA},
except for $^{24}$O (Fig. \ref{fig:ormsr}).
The reason for the failure in reproducing the enhancement of the radius
near to the drip line is conjectured that the present model space 
is not sufficient to describe the detailed behaviour of 
weakly bound neutrons.

\subsection{nuclear matter property}

\begin{figure}
\noindent
\epsfxsize=0.4\textwidth
\centerline{\epsffile{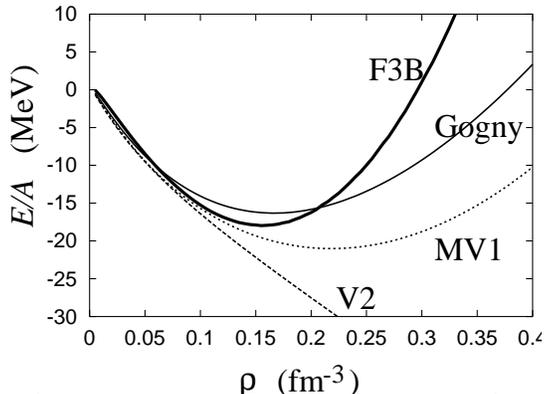}}
\caption{\label{fig:matter}
Equation of state for symmetric nuclear matter.
The energy per nucleon is plotted as a function of the density.
The effective forces are Gogny D1, F3B, Volkov No.2($m=0.65$), and MV1
case 1($m=0.60$). 
}
\end{figure}

The matter properties were evaluated by plane waves in the Fermi sea.
The equation of state(EOS) of the symmetric nuclear matter
is illustrated in Fig. \ref{fig:matter}.
The properties at the saturation point
are listed in Table \ref{tab:matter}.
The SIII force, Gogny D1 force, Tohsaki's F1 force, F3B force
give similar results for the saturation energy $E_s/A$ and 
the saturation density $\rho_s$($\rho_s=2 k_f^3/3\pi^2$, where $k_f$ is Felmi
momentum at the saturation point).
On the other hand, the MV1 force can not satisfactorily explain
the following empirical values: $E_s/A \sim -16$ MeV and $\rho_s \sim 0.16$
fm$^{-3}$($k_f \sim 1.33$ fm$^{-1}$).

The small incompressibility of matter($K_\infty$), namely, the
soft-type equation of state(EOS) is given by such effective forces as the 
Gogny force and the SLy4 forces(a Skyrme parametrization)\cite{CHABANAT}, 
while the F3B, SIII, Tohsaki's F1, MV1 forces give large values of 
incompressibility, which correspond to the hard EOS 
(see table \ref{tab:matter}).
Comparing EOS for the F3B force with that for the Gogny D1 force, 
it is fount that
the EOS is similar to each other in the low-density region, while
the matter energy for the F3B force rapidly increases in the high-density region.
The incompressibility of nuclear matter has been discussed
in relation to the energy of giant monopole resonances(GMR)
\cite{BLAIZOT,COLO,YOUGBLOOD,LALAZISSIS}.
Although a soft-type equation of state(EOS), like $K_\infty=200\sim 240$,
is preferred to quantitatively describe the GMR energy
in HF analysis with Gogny and Skyrm-type effective forces \cite{COLO,BLAIZOT}, 
there still remains an inconsistency with the theoretical values 
suggested by a relativistic-mean-field analysis \cite{LALAZISSIS},
where larger incompressibilities are predicted. On the other hand,
the hard-type Tohsaki's F1 force reasonably reproduces the 
GMR energy of $^{40}$Ca \cite{TOHSAKInew}.

\begin{figure}
\noindent
\epsfxsize=0.4\textwidth
\centerline{\epsffile{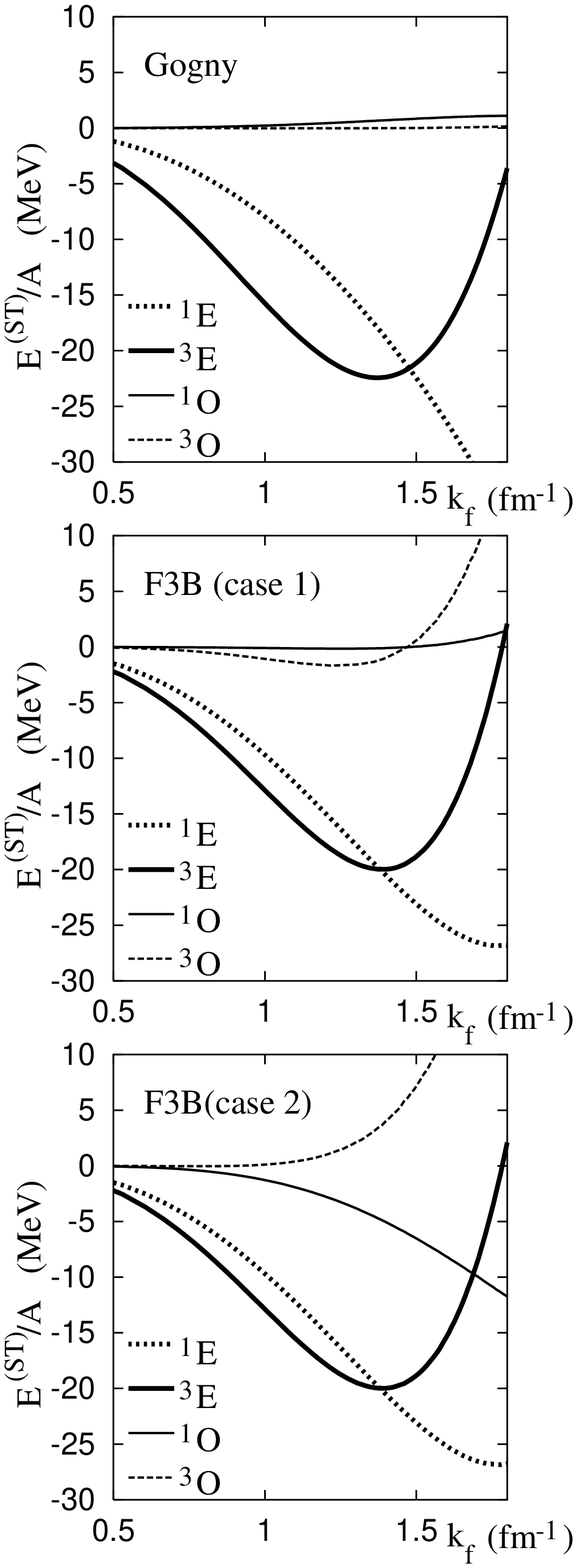}}
\caption{\label{fig:eos-part}
Partial contributions $E^{(ST)}$ 
of the potential energy in each subspace of 
singlet even($^1$E),
triplet even($^3$E), singlet odd($^1$O) and triplet odd($^3$O)
defined by the projectors $P^{(ST)}=(1\pm P_\sigma)(1\pm P_\tau)/4$,
where (ST)=(0,1), (1,0), (0,0) and (1,1), respectively. 
The calculated values of the partial potential energy per nucleon are   
plotted as functions of the Fermi momentum($k_f$).}
\end{figure}

It has been known that matter saturation
can not be described by effective two-body central forces 
without any repulsive density-dependent or three-body terms,
as can be seen in the EOS given by the Volkov force (Fig. \ref{fig:matter}).
In other words, the density-dependent or three-body terms are essential
to explain the nuclear matter saturation.
In order to exhibit the mechanism of nuclear saturation,
we decomposed the total potential energy into different contributions,
$E^{(ST)}$, in each subspace $(ST)$, defined
by the projectors $P^{(ST)}=(1\pm P_\sigma)(1\pm P_\tau)/4$ in the 
same way as is the analysis described in Ref.\cite{GOGNY}.
The decomposed potential energy of the F3B forces 
is shown in Fig. \ref{fig:eos-part}.
The dominant parts of the potential energy at a saturation density
of $k_f \sim 1.3$ fm$^{-1}$ are  
the singlet even($^1 E$) and triplet even($^3 E$) terms, while the 
singlet odd ($^1$O) and triplet odd ($^3$O) parts
are minor. We notice that the $^3 E$ part plays an 
important role in the saturation mechanism.
This result is consistent 
with a tendency which was indicated in the nuclear matter calculations
by Brueckner-Goldstone Theory using realistic nucleon-nucleon interactions 
 \cite{BETHE}.
It is very reasonable because the matter saturation is considered 
to come from a suppression of the tensor part of the 
bare nucleon-nucleon interaction 
in nuclear matter,
because the Pauli principle acts to reduce the attractive tensor force 
due to a blocking of the intermediate states.
Therefore, tensor suppression in nuclear matter is regarded to be 
phenomenologically simulated by the repulsive density-dependent
terms in the effective central forces. In that sense, the repulsive
density-dependent term should be dominated by the triplet even component,
just as in the case of the tensor force. As a result, 
the triplet even component indicates the saturation behavior with an
increase of density.
Strictly speaking, also, the $^3$O part of the potential energy 
has an enhancement in the region $k_f \ge 1.5$ fm$^{-1}$,
because the three-body repulsive term is not pure $^3E$, but it also contains 
the $^3$O component
because of the simple form of the three-body operator 
in F3B forces.
Such an enhancement in a rather high-density region may have
at least no significant effect 
on the structure of finite nuclei.

\begin{table}
\caption{ \label{tab:matter} Saturation Properties 
for nuclear matter. The saturation energy, Felmi momentum, and 
incompressibility at the saturation point for various effective forces.}
\begin{center}
\begin{tabular}{cccccccc}
 Force  & MV1(case1 m=0.60) & SIII & SLy4 & Gogny(D1) & Tohsaki's F1 & F3B(this work)\\
\hline
 $E/A$(MeV) & $-$21.01 & $-$15.87 & $-$15.97& $-$16.32& $-$17.0 & $-$17.9\\
 $k_f$(fm$^{-1}$) & 1.48 & 1.29 & 1.33 & 1.35& 1.27 & 1.32 \\
 $K_\infty$ (MeV) & 299 & 356 & 230 & 228 & 309  & 390 
\end{tabular}
\end{center}
\end{table}

\section{Discussion}\label{sec:discuss}

\subsection{Significance of finite-range three-body term}

\begin{figure}
\noindent
\epsfxsize=0.4\textwidth
\centerline{\epsffile{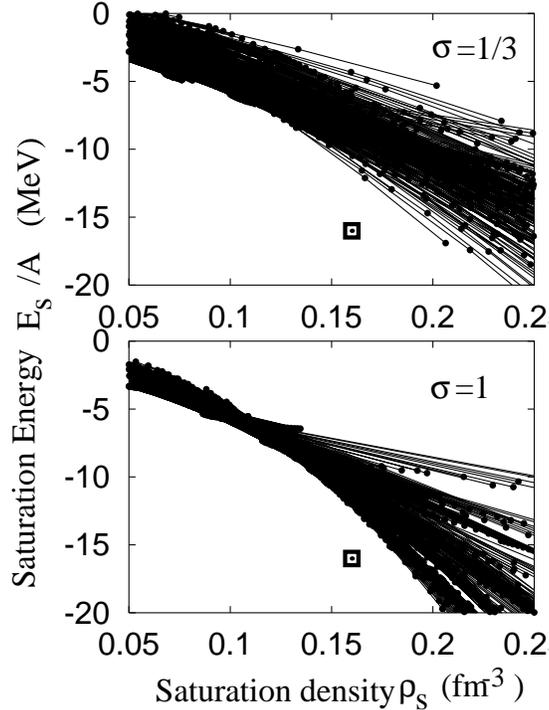}}
\caption{\label{fig:alpha-matter}
Dots indicate the saturation energy and density in symmetric nuclear matter  
obtained by effective central forces with 
finite-range two-body terms and a zero-range density-dependent term,
which satisfy the experimental data of the size and energy of 
an $\alpha$ particle.
The square points correspond to the empirical saturation energy and 
density ($\rho_s=0.16$ fm$^{-3}$, $E_s/A=-16$ MeV). 
The details are explained in the text.
}
\end{figure}

The finite-range three-body terms in the phenomenological effective 
forces are important to describe the overall properties 
for nuclei for a broad range of masses.
As mentioned in \ref{sec:results}, and also 
already argued in the pioneering work by Tohsaki \cite{TOHSAKInew},
the finite-range three-body terms are empirically 
essential to simultaneously reproduce the properties of 
$\alpha$-particle(size and $\alpha$-$\alpha$ phase shift) 
and nuclear matter saturation.
Moreover, the finite-range feature is reasonable considering the original
range of the tensor force, whose suppression effect in matter 
is simulated by the three-body terms.
In order to explain the indispensability of the finite-range 
nature in three-body terms,
we demonstrate the fatal failure of general effective forces
with zero-range density-dependent terms
in describing the correct size and energy of an $\alpha$ particle and 
the saturation density and energy of nuclear matter.
We assume an effective central force, which consists of 
a two-body term with a two-range Gaussian form and 
a zero-range density-dependent term, as follows:
\begin{eqnarray}
& V_{ij}=(1-m-mP_\sigma P_\tau)\left[v_1 \exp(-\frac{r_{ij}^2}{a_1^2})
+v_2 \exp(-\frac{r_{ij}^2}{a_2^2})\right ]+ v_3 \rho^\sigma(r_{ij})\cdot
 \delta(r_{ij}),\\
{\rm where} & \sigma=\frac{1}{3} \ {\rm or}\ 1.
\end{eqnarray}
Coulomb force is omitted.
By assuming the $(0s)^4$ HO wave function for an $\alpha$ particle and
plane waves in the Fermi sea for nuclear matter,
we can easily obtain the optimum size($r_\alpha$) and 
energy($E_\alpha$) of an $\alpha$ particle 
and the matter saturation density($\rho_s$) and energy($E_s$) 
as a function of the parameter set ($v_1, a_1, v_2, a_2, v_3, m$) as
$r_\alpha(v_1, a_1, v_2, a_2, v_3, m)$ ,
$E_\alpha(v_1, a_1, v_2, a_2, v_3, m)$ ,
$\rho_s(v_1, a_1, v_2, a_2, v_3, m)$ , and 
$E_s(v_1, a_1, v_2, a_2, v_3, m)/A$. 
We try to search for a parameter set which can reproduce the empirical data
of these four values.
Firstly, we put the following conditions: that 
the size and energy of an $\alpha$ particle are equal to
the experimental data,
$r_\alpha =1.48$ fm and $E_\alpha =29.253$ MeV.
Here, the experimental input of the energy is a modified one
with a Coulomb-force correction. Keeping these two constraints, we
vary the parameters ($v_1, a_1, v_2, a_2, v_3, m$) in the 
following ranges:
$0.6 \le a_1 \le 2.5$ fm,
$0.5 \le a_2 \le a_1$ fm,
$100 \le v_3 \le 4000$ MeV fm$^4$($\sigma=1/3$) or  
$100 \le v_3 \le 4000$ MeV fm$^6$($\sigma=1$),
$0.01 \le m \le 0.99$, and investigate
the behaviour of the matter saturation point $(\rho_s, E_s/A)$,
given by the various sets
of paramters.
We also put a physical restriction, $v_1 \le 0$, that indicates that 
the long-range term must be attractive.
The saturation points $(\rho_s, E_s/A)$ are 
displayed in Fig. \ref{fig:alpha-matter}. The upper panel indicates 
the saturation points in the case $\sigma=1/3$, and
the lower panel corresponds to the case $\sigma=1$.
Under the conditions of the
 $\alpha$-particle properties, the saturation points 
can not reach the empirical saturation point
($\rho_s\sim 0.16$ fm$^{-3}$, $E_s/A \sim -16$ MeV)
in both cases, because the allowed area of the saturation points 
does not include the empirical one.
This means that there is no reasonable parameter 
set which is able to simultaneously describe the saturation properties 
of an $\alpha$ particle and nuclear matter.
In other words, 
with ``zero-range'' density-dependent-type
effective central forces, which explain the 
matter saturation point,
it is difficult to describe the compact size of a real 
$\alpha$ particle.
From these results, we can conclude that the finite-range terms are 
indispensable to simultaneously reproduce the properties of 
$\alpha$-particle and matter saturation.

\subsection{Relation with $s$-wave nucleon-nucleon scattering}

It has already been known that 
bare nucleon-nucleon interactions have leading tensor terms
and hard cores, both of which are not easily treated in the 
microscopic calculations of nuclear many-body systems.
Instead, one usually adopts phenomenological effective nucleon-nucleon 
forces dominated by central forces with no hard cores.
It is regarded that the tensor terms and hard cores are renormalized in the 
effective central forces.
As already mentioned, the necessity of repulsive 
density-dependent terms of effective forces for describing 
the saturation properties
originates from a suppression of the attractive tensor force 
in nuclear matter\cite{BETHE}.
In other words, the density dependence of the tensor force suppression
is phenomenologically simulated by density-dependent repulsive terms  
in the effective central forces.
From this point of view, we demonstrate a link between the effective
forces and nucleon-nucleon(N-N) scattering and discuss the role of the
density-dependent repulsive terms.
Firstly, we reproduce the $s$-wave N-N scattering phase 
shift by the central two-body forces in the F3B force. Second,
we discuss the role of the residual part with the three-body term,
which should be connected with the matter effect on the N-N
interaction.

\begin{figure}
\noindent
\epsfxsize=0.4\textwidth
\centerline{\epsffile{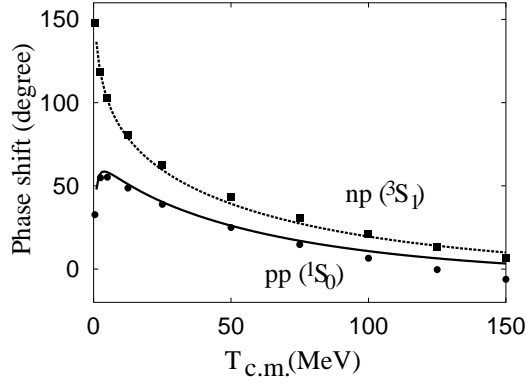}}
\caption{\label{fig:nn}
$s$-wave phase shift of nucleon-nucleon scattering calculated
with the effective two-body central force $V^{{\rm 2-body}}$
defined in equations \protect\ref{eqn:v2-1}-\protect\ref{eqn:v2-4} 
of the text.}
\end{figure}

\begin{figure}
\noindent
\epsfxsize=0.4\textwidth
\centerline{\epsffile{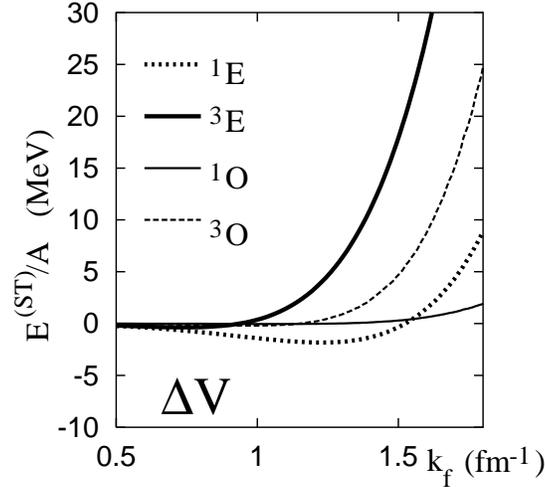}}
\caption{\label{fig:eos-del}
Partial contributions $E^{(ST)}$ 
of the nuclear matter potential energy for $\Delta V$ in each subspace 
defined by the projectors $P^{(ST)}=(1\pm P_\sigma)(1\pm P_\tau)/4$.
}
\end{figure}

We decompose the central part of the present effective force(F3B) into 
two terms.
One is a two-body term, $V^{\rm 2-body}$, which can reproduce the $s$-wave 
nucleon-nucleon scattering behaviour,
and the other is the residual term, $\Delta V$, which contains the three-body
term, defined as follows:
\begin{eqnarray}
& V_{central}=\label{eqn:v2-1} \sum_{i <j } V^{(2)}_{ij}+\sum_{i<j<k} V^{(3)}_{ijk} =
V^{{\rm 2-body}}+\Delta V,\label{eqn:decompose}\\
& V^{{\rm 2-body}}= \sum_{i <j } V^{(2)}_{ij} + \sum_{i <j } V'_{ij},\\
& \Delta V= -\sum_{i <j } V'_{ij}+\sum_{i<j<k} V^{(3)}_{ijk} \\,
& \label{eqn:v2-4} 
V'_{ij}=v' \exp\left[-(\frac{{\bf r}_i-{\bf r}_j}{a^{(2)}_3})^2 \right], 
\end{eqnarray}
where $v'=32.5$ MeV. Other interaction parameters in $V_{ij}^{(2)}$ 
and $V_{ijk}^{(3)}$ are same as those of 
F3B force.
The first term $V^{{\rm 2-body}}$ in Eq. \ref{eqn:decompose}
is the two-body central force.
The $s$-wave phase shift for the low-energy N-N
scattering data
is well described by this two-body term $V^{{\rm 2-body}}$,
as shown in Fig. \ref{fig:nn}.
Moreover, the calculated scattering lengths of the singlet and 
triplet $s$-wave for $V^{{\rm 2-body}}$
are $a_s=5.37$(fm) and $a_t=-19.6$(fm),
which reasonably agree with the experimental 
values, $a_s=5.39$ and $a_t=-23.7$. 
Therefore, we consider this $V^{{\rm 2-body}}$ as an effective
force between bare nucleons, where the tensor force in the realistic 
interaction is renormalized in the central term.
Thus, the second term($\Delta V$) 
in Eq. \ref{eqn:decompose} can be understood in terms of the matter effect
in finite nuclei and nuclear matter.
Since the leading matter effect originates from 
a suppression of the tensor force, $\Delta V$ should be repulsive, 
and behaves as triplet even.
In Fig. \ref{fig:eos-del}, we represent the partial contributions,
$E^{(ST)}$, of the potential energy given by $\Delta V$
in each subspace $(ST)$ of nuclear matter.
At a saturation density of $k_f\sim 1.3$ fm$^{-1}$,
the dominant component of $\Delta V$ is the triplet even part. 
This means that the density-dependent repulsive term($\Delta V$)
actually takes the role of tensor force suppression
in nuclear matter.

As mentioned above, the central effective force(F3B) is decomposed into
two parts, $V^{\rm 2-body}$ and $\Delta V$. In principle, it is easy to 
understand the decomposition if $V^{\rm 2-body}$ is same as the two-body term 
$\sum_{i <j }V^{(2)}_{ij}$ of the F3B force, and $\Delta V$ consists of only
the three-body term of F3B. However, in the present analysis, 
$V^{\rm 2-body}$ is the modified one
from the original $\sum_{i <j } V^{(2)}_{ij}$, and $\Delta V$ involves
a two-body term in addition to the dominant three-body term.
We consider the reason as following way. The origin of the finite-range 
three-body term may not be the pure three-body effect but it also 
contains some component from the two-body central force.
It is because of the restriction of the simple finite-range 
three-body operator,
where we adopt a common range for the 
distances of all nucleon pairs in three-body term 
and omit spin-dependent terms for simplicity
to apply to practical calculations with AMD method. 
In order to overcome the restriction,
we consider that $\Delta V$, 
which consists of the two-body and thee-body terms,
indicates the matter effect and means the effective 
density-dependent repulsive force.

\section{Summary}\label{sec:summary}

We proposed new effective inter-nucleon forces with a finite-range
three-body operator for microscopic calculations of nuclear structure.
It was found that the proposed forces are suitable 
to describe saturation properties over a
wide mass number region from the $\alpha$ particle to 
nuclear matter. Compared with similar-type
forces proposed by Tohsaki, 
an advantage of the new forces is that they are more practical 
because of the simple three-body operator, and therefore are applicable 
to microscopic calculations of general nuclei.
The proposed effective forces were applied to 
AMD+GCM calculations of $Z=N$ ($A\le 40$) nuclei and O isotopes.
In addition to the success in systematically reproducing the
 radii and energy of these nuclei, we demonstrated the
characteristics of the new forces concerning the $\alpha$-$\alpha$
scattering behaviour and matter EOS.
Especially, the new forces successfully describe the 
compact size of a real $\alpha$ particle and the $\alpha$-$\alpha$ 
scattering phase shift, which typical effective forces with
zero-range density-dependent terms can not reproduce.
Concerning the matter properties, we found that the saturation mechanism 
in the nuclear matter originates from the dominant triplet even component 
in the repulsive three-body term. 
The new forces were also discussed in relation with 
the $s$-wave N-N scattering behavior.

The density-dependent repulsive terms are essential to describe 
nuclear saturation. The origin is a suppression of the tensor force of 
bare nucleon-nucleon interactions in nuclear matter.
From this point of view, we discussed the role of the 
finite-range three-body term in the new forces.
The importance of the finite-range three-body term was also described 
based on empirical concepts.
We conclude that the finite-range three-body operator plays an essential
role in describing the overall properties over a wide mass number region
as well as the matter saturation properties.

\acknowledgments
The authors would like to thank Dr. A. Ono, Prof. A. Tohsaki and 
Prof. H. Horiuchi and Prof. K. Ikeda for helpful discussions and comments.
They are thankful to
Dr. O. Morimatsu and Dr. H. Nemura for technical advise. 
A part of the calculations were performed with the AMD code(version 2.5)
provided by Dr. A. Ono.
The computational calculations of this work were supported by 
the Supercomputer Project No.58, No.70 and No.83 of 
High Energy Accelerator Research
Organization(KEK), and Research Center for Nuclear Physics 
in Osaka University.
This work was supported by Japan Society for the Promotion of 
Science and a Grant-in-Aid for Scientific Research of the Japan
Ministry of Education, Science and Culture.
The work was partially performed in the ``Research Project for Study of
Unstable Nuclei from Nuclear Cluster Aspects'' sponsored by
Institute of Physical and Chemical Research (RIKEN).



\begin{thebibliography}{99}
  
\bibitem{AMDrev}
Y. Kanada-En'yo, M. Kimura and H. Horiuchi, Comptes rendus Physique Vol.4, 497(2003).

\bibitem{ENYObc}
 Y. Kanada-En'yo, H. Horiuchi and  A. Ono,
Phys. Rev. C {\bf 52}, 628 (1995);
 Y. Kanada-En'yo and H. Horiuchi,
Phys. Rev. C {\bf 52}, 647 (1995).

\bibitem{ENYOsup}
Y. Kanada-En'yo and  H. Horiuchi, Prog. Theor. Phys. Suppl.{\bf 142},
 205(2001).
\bibitem{BONCHE}
P. Bonche et al., Nucl. Phys. {\bf A443}, 39 (1985).
\bibitem{TAJIMA}
N.Tajima, S.Takahara and N.Onishi,
Nucl.Phys. {\bf A603}, 23 (1996).
\bibitem{TAKAMI}
S.Takami, K.Yabana and K.Ikeda,
Prog.Theor.Phys. {\bf  96}, 407 (1996).
\bibitem{MINNESOTA}
D. R. Thompson, M. Lemere and Y. C. Tang,
Nucl. Phys. {\bf 286}, 53(1977).
\bibitem{VOLKOV} A. B. Volkov, Nucl. Phys {\bf 74}, 33 (1965).
\bibitem{GOGNY} J. Decharge and D. Gogny, Phys. Rev. {\bf C21} 1568,
        (1980).
\bibitem{SIII} M. Beiner, H. Flocard, Nguyen van giai and P. Quentin,
        Nucl. Phys. {\bf A 238} (1975), 29.
\bibitem{CHABANAT}
E. Chabanat, P. Bonche, P. Haensel, J. Meyer, R. Schaeffer,
Nucl. Phys. {\bf A635}, 231 (1998).
\bibitem{MVOLKOV} T. Ando, K. Ikeda and A. Tohsaki,
        Prog. Theory. Phys. {\bf 64}, 1608  (1980).
\bibitem{TOHSAKInew}
A. Tohsaki, 
Phys. Rev. {\bf C49}, 1814 (1994).
\bibitem{KAMIMURArgm}
M. Kamimura, Prog. Theor. Phys. Suppl. No. 62, 236 (1978).
\bibitem{OKABE-be9}
S. Okabe and Y. Abe, Prog. Theor. Phys. {\bf 61}, 1049 (1979).
\bibitem{aa-phase}
R. Nilson, W. K. Jentschke, G. R. Briggs, R. O. Kerman and J. N. Snyder,
Phys. Rev. {\bf 109}, 850 (1958);
T. A. Tombrello and L. S. Senhouse, Phys. Rev. {\bf 129}, 2252 (1963);
N. P. Heydenberg and G. M. Temmer, Phys. Rev. {\bf 104}, 123 (1956);
W. S. Chien and R. E. Brown, Phys. Rev. {\bf C10}, 1767 (1974). 
\bibitem{ENYOe}
 Y. Kanada-En'yo,
Phys. Rev. Lett. {\bf 81}, 5291 (1998).
\bibitem{G3RS}
N. Yamaguchi, T. Kasahara, S. Nagata and Y. Akaishi,
Porg. Theor. Phys. {\bf 62}, 1018 (1979).
\bibitem{SUGAWA}
Y. Sugawa, M. Kimura and H.Horiuchi,
Prog. Theor. Phys. {\bf 106}, 1129 (2001).
\bibitem{NDATA}
G. Fricke, C. Bernhardt, K. Heilig, L. A. Schaller, L. Schellenberg, 
E. B. Shera and C. W. Dejager, 
Atomic Data and Nucl. Data Tabl. {\bf 60}, 177 (1995).
\bibitem{OZAWA}
A.Ozawa, T.Suzuki, I.Tanihata, 
Nucl. Phys. {\bf A693}, 32 (2001).

\bibitem{BLAIZOT}
J. P. Blaizot, Phys. Rep. {\bf 64}. 171(1980);
J. P. Blaizot et al., Nucl.Phys. {\bf A591}, 435 (1995).

\bibitem{COLO}
G. Col\`o, P. F. Bortignon, N. Van Gai, A. Bracco, and
R. A. Broglia, Phys. Lett. B {\bf 276}, 279(1992).
\bibitem{YOUGBLOOD}
D. H. Youngblood, H. L. Clark, and Y. -W. Lui,
Phys. Rev. Lett. {\bf 82}, 691 (1999).

\bibitem{LALAZISSIS}
G. A. Lalazissis, J. K\"onig, and P. Ring, Phys. Rev. C{\bf 55}, 540
(1997). 

\bibitem{BETHE}
H. A. Bethe, Ann. Rev. Nucl. Sci. {\bf 23}, 93 (1971).


\end{thebibliography}
\end{document}